\documentclass[twocolumn,twoside,conference]{IEEEtran}
 \usepackage{multirow}
\usepackage[dvips]{graphicx}
\usepackage[english]{babel}
\usepackage[latin1]{inputenc}
\usepackage{amsfonts,amssymb,amsmath,dsfont,amscd,url,graphics,cite,calc,psfrag,theorem, epsfig}
\usepackage{url}
\usepackage{cite}
\usepackage{stfloats}
\usepackage{graphicx}
\usepackage[dvipsnames]{xcolor}
\usepackage{rotating}
\usepackage{tikz}
\usepackage{pgf}
\usepackage{pgffor}
\usepackage{pgfpages}
\usepackage{wrapfig}
\usepackage{algorithm}
\usepackage{algorithmic}
\usepackage{datetime}
\ifx\pdftexversion\@undefined
  \usepackage[pdftex]{graphicx}
\else
  \usepackage[dvips]{graphicx}
\fi
 \usetikzlibrary{arrows}
 \usetikzlibrary{snakes}
 \usetikzlibrary{shapes}
 \usetikzlibrary{automata}
 \usetikzlibrary{topaths} 
\usetikzlibrary[topaths] 
\usetikzlibrary{positioning}
\usepgfmodule{decorations}
\usetikzlibrary{decorations.pathmorphing}
\pgfdeclarelayer{background} \pgfdeclarelayer{foreground} %
\pgfsetlayers{background,main,foreground}
\usetikzlibrary{calc,3d}
\usetikzlibrary{shapes.gates.logic.US,trees,positioning,arrows}
\usetikzlibrary{positioning}
\usetikzlibrary{decorations.pathmorphing} 
\usetikzlibrary{fit}	
\usetikzlibrary{shapes,snakes}
 \usetikzlibrary{arrows}
 \usetikzlibrary{snakes}
 \usetikzlibrary{shapes}
 \usetikzlibrary{automata}
 \usetikzlibrary{topaths} 
\usetikzlibrary[topaths] 
\pgfdeclarelayer{background} \pgfdeclarelayer{foreground} %
\pgfsetlayers{background,main,foreground}

\par
\floatstyle{ruled}
\newfloat{alg}{htb}{loa}
\floatname{alg}{Algorithm}

\floatstyle{plain}
\newfloat{alg1}{htb}{loa}
\floatname{alg1}{Algorithm}
\usepackage{wrapfig}\usepackage{subfig}
\newif\iffinal
\usepackage{tikz}
\colorlet{drkblue}{blue!80}
\iffinal
  
\else
  
\fi
\makeatother

\newtheorem{Proposition}{Proposition}

\newtheorem{definition}{Definition}

\newtheorem{rmk}{Remark}

\usepackage{multicol}
\usepackage{trsym,trfsigns}
\usepackage{lipsum}
\usepackage{pdfpages}
\usepackage{epstopdf}
 \graphicspath{{Figs/}}
\pgfrealjobname{COSPA_JnlSubmitArXiv} 
\begin{document}
\title{A New Optimal Subpattern Assignment (OSPA) Metric  for Multi-target Filtering
	\author{\IEEEauthorblockN{Tuyet~Vu}
		\IEEEauthorblockA{
			Intelligence, Surveillance and Space Division\\
			Defence Science and Technology Group, Australia\\
			Email:Tuyet.Vu@defence.gov.au}
	}
}
\pagestyle{empty}

\maketitle
\begin{abstract}
This paper proposes and evaluates a new metric.
This metric will overcome a limitation of the Optimal Subpattern
Assignment (OSPA) metric mentioned by Schuhmacher et al.: the
OSPA distance between two sets of points is insensitive to the
the case where one is empty. This proposed metric called Complete
OSPA (COSPA), retains all the advantages of the OSPA metric for
evaluating the performance of multiple target filtering algorithms
while also allowing separate control over the threshold of physical
distance errors and cardinality errors.

\end{abstract}
\selectlanguage{english}
\IEEEpeerreviewmaketitle
\section{Introduction}\label{S:intro}
In the area of multi-target filtering (MTF), a rigorous and robust metric plays an important role for performance evaluation. The application of metric can be found in target tracking such as . A key technical issue underpinning such an evaluation concerns the measurement of `distance' between two sets. Clearly a standard distance measure on Euclidean space cannot be used. Hoffman and Mahler \cite{HoffmanMahler02} studied this problem and found that the Hausdorff distance is relatively insensitive to errors in the number of targets (which is an important issue in MTF) and proposed a new metric to overcome this shortcoming. In $2008$, Schuhmacher et al. \cite{Schuhmacher2008} built on this work and proposed a metric called optimal sub-pattern assignment (OSPA) which incorporated the spatial distance and the cardinality distance. Spatial distances between pairs of state vectors across two sets are cut off at $c$, a parameter, and are weighted equally while each extra element in a bigger set will be penalized as if there was a distance error of $c$. However, when one set is empty, the OSPA distance takes on the value of $c$ regardless of the cardinality of the other set. This was pointed out in \cite{Schuhmacher2008} as a minor inconvenience. However, as discussed in Rahmathullah et al. \cite{rah2016GOSPA
}, in reality the OSPA metric is not a desirable tool for evaluating MTF algorithms. They pointed out that the MTF community prefers to understand the missing target and false target performance beyond the cardinality mismatch. Hence Rahmathullah et al. \cite{rah2016GOSPA} tried to overcome these limitations by proposing a new metric called Generalized OSPA (GOSPA). This metric was derived by removing the normalization of the OSPA metric and multiplying parameter $1/\alpha$ to the cardinality error with the optimal choice of $\alpha=2$. However, by not normalizing, the GOSPA metric will generally grow with the size of the sets. 
The normalization of the OSPA metric scales the sum of all distances between the two finite sets to be within the interval $[0,c]$. In contrast, the GOSPA metric is exactly the sum of all distances between the finite sets. 
Furthermore, by weighting the cardinality penalty less than the spatial distance cut-off $c$, the GOSPA metric gives unexpected results with most scenarios when the sets are not empty (see example in Figure \ref{Fig:Ex_RemDe}). Another issue is that the OSPA metric is insensitive to the distance  cut-off and cardinality penalty.  
The cause is that the penalty for each unassignable point and the distance cut-off between two assigned points in the OSPA metric are the same, $c$.

By analysing the advantages and disadvantages of OSPA together with the requirements of the MTF community as discussed in \cite{rah2016GOSPA}, we propose a new metric, namely a Complete OSPA (COSPA). This metric overcomes the shortcoming of the OSPA metric when one of the two finite sets is empty. Also, by choosing the cut-off value to be less  than or equal to the penalty for each unassignable point, together with considering the sensitivity to the empty set, the COSPA metric overcomes the limitations of the OSPA metric while still retaining all the advantages of the OSPA metric for other cases when two finite sets are non-empty. The earlier result is published in \cite{Tuyet2020}. 

The structure of this paper is as follows. Section \ref{S:Background} provides some definitions and background. Section \ref{S:OSPA} summarizes the OSPA metric and discusses its limitations while
Section \ref{S:GOSPA} summarizes and discusses the GOSPA metric. The main part of this paper is
Section \ref{S:COSPA_metric}, where the COSPA metric is developed and the analysis of OSPA, GOSPA and COSPA is conducted,  with illustrative scenarios discussed in Section \ref{S:Ana-COSPA}. 
Section \ref{S:Experiment} presents numerical studies to demonstrate the usefulness of the proposed metric and compares this with the other two metrics mentioned above. 
Concluding remarks and suggestions for future work are given in Section \ref{S:Conclusion}.
\section{Background\label{S:Background}}
Some background and definitions in this section are used throughout the paper.
Let $\mathcal{X}\subseteq\mathds{R}^{n_x}$ 
where $n_x$ is the natural number. The distance $\bar{d}^{(a)}
(x,y)$ \footnote{An arbitrary function with values in $[0,a]$ could be used. 
	In particular, $\bar{d}^{(a)}
	(x_i,y_{\pi(i)})=t(d(x,y))$ where $t:[0,\infty)\to [0,a]$ is any transformation that is nondecreasing, subadditive and $t(u)=0$ if and only if $u=0$.} for $x,y\in\mathcal{X}$ and $a>0$ is cut off by $a$ if it is equal to or larger than $a$. i.e.
\begin{align}\label{E:dinstance_c}
	\bar{d}^{(a)}
	(x,y)=\min (a,d(x,y))
\end{align}
where $d(x,y)$ is typically $\|x-y\|_{p'}$ where $1\le p'\le \infty$. 

A finite set is a collection of a finite number of elements. If $A$ is a finite set, then $A$ has finite number of elements, i.e. $|A|<\infty$. A finite set $B$ is defined to be bigger than a finite set $A$ if $B$ has more elements than $A$, i.e. $|B|>|A|$.

Let $A$ and $B$ be two finite sets. Denote $\Pi(A,B)$ as a set of all one-to-one functions between $A$ and $B$, i.e. if $|A|\le|B|$ then $\pi\in\Pi(A,B)$ is a one-to-one (injective) function from $A$ to $B$ otherwise 
from $B$ to $A$. Mathematically, $\Pi(A,B)$ is
\begin{align}\label{E:Permutation}
	\hspace{-0.2cm}\left\{\!\!
	\begin{array}{ll}
		\{\pi\text{ from }A\text{ to }B:\pi \text{ is injective}\}
		,\!\!\!&\hbox{if $|A|\le|B|$;} \\
		\{\pi\text{ from }B\text{ to } A:\pi \text{ is injective}\},&\hbox{otherwise.}
	\end{array}
	\right.\!\!
\end{align}
\section{OSPA\label{S:OSPA}}
We summarize the OSPA metric in Section \ref{SS:OSPA} and discuss its limitations in Section \ref{SS:LimitOSPA}.
\subsection{Summary\label{SS:OSPA}}
 The OSPA distance, the total error, is the sum of a localization error and a cardinality error. The localization error is the smallest sum of distances between all combinations of elements of two finite sets $X$ and $Y$. The cardinality error is given by the mismatch between the number of elements in $X$ and the number in $Y$ scaled using $c$. 
The per target error obtained by normalizing total error by the largest cardinality of the two given sets is a proper metric.
\begin{definition}[OSPA metric]\label{D:OSPA}
	Let $X$ and $Y$ be two finite sets. For order parameter $p$ ($p\geq 1$), the cut-off parameter $c$ ($c>0$), the OSPA metric $\bar{d}^{(c)}_p(X,Y)=\bar{d}^{(c)}_p(Y,X)$ is defined as.
	\begin{itemize}
		\item if $X=Y=\emptyset$: $\bar{d}^{(c)}_p(X,Y)=0.$
		\item otherwise if $p<\infty$, $\bar{d}^{(c)}_p(X,Y)$ is for $|X|\le|Y|$
		\begin{align}\label{E:OSPA}
			\hspace{-0.7cm}
			\left[\frac{1}{|Y|}\left(\!\displaystyle\min_{\pi\in\Pi(X,Y)}\!\sum_{x\in X}\!\bar{d}^{(c)}
			(x,\pi(x))^p\!+\!c^p(|Y|-|X|)\!\right)\!\right]^{\frac{1}{p}}\!\!\!\!\!
		\end{align}
		\item if $p=\infty$, $\bar{d}^{(c)}_p(X,Y)$ is
		\begin{align}\label{E:OSPA_infty}
			\left\{
			\begin{array}{ll}
				\displaystyle\min_{\pi\in\Pi(X,Y)}\max_{x\in X}\bar{d}^{(c)}
				(x,\pi(x)),&\hbox{if $|X|=|Y|$;} \\
				c&\hbox{if $|X|\ne |Y|$;}
			\end{array}
			\right.
		\end{align}
	\end{itemize}
	where 
	$\Pi(X,Y)$ is 
	in \eqref{E:Permutation}; and the choice of $c$ and the steps required to calculate the OSPA metric can be found in \cite{Schuhmacher2008}. 
\end{definition}
The OSPA distance is interpreted as a $p-$th order per-target error, comprised of a $p-$th order per-target localization error and a $p-$th order per-target cardinality error. 
By \eqref{E:OSPA} for $1\le p<\infty$ , the localization and cardinality errors are given respectively as follows for $|X|\le|Y|$
\begin{align*}
	\bar{e}^{(c)}_{p,\text{loc}}(X,Y)&=\bar{e}^{(c)}_{p,\text{loc}}(Y,X)\\&=\left(\frac{1}{|Y|}\min_{\pi\in\Pi(X,Y)}\sum_{x\in X}\!\bar{d}^{(c)}
	(x,\pi(x))^p\right)^{\frac{1}{p}}\!\!\!\!\!\!,\nonumber\\
	\bar{e}^{(c)}_{p,\text{card}}(X,Y)&=\bar{e}^{(c)}_{p,\text{card}}(Y,X)=c\left(\frac{|Y|-|X|}{|Y|}\right)^{\frac{1}{p}}\!\!\!\!\!\!.
\end{align*}
The decomposition of the OSPA metric into separate components is usually not necessary for performance evaluation, but may provide valuable additional information.
\subsection{Discussion of the OSPA metric}\label{SS:LimitOSPA}
By Definition \ref{D:OSPA}, if $X=\emptyset$ and $Y\ne\emptyset$, $\bar{d}^{(c)}_p(X,Y)=c$ regardless of how many elements there are in set $Y$ (see example in Figure \ref{Fig:Ex_CutoffDifPenCarAlphaSm1} and the discussion in Section \ref{S:Ana-COSPA}). 

Furthermore, the OSPA metric gives the same result for the distance between two sets $X$ and $Y$ where $|Y|=|X|-1$, and between two sets $X$ and $Y\cup\{z\}$ where $d(x,z)\ge c$ for $z\notin Y$ and $\forall x\in X$ (see example in Figure \ref{Fig:ExSameExtraPoint}). The GOSPA metric \cite{rah2016GOSPA} overcomes these limitations but it has shortcomings compared with OSPA. The next section summarizes and discusses the GOSPA metric.
\section{Generalized OSPA (GOSPA)}\label{S:GOSPA}
The GOSPA metric \cite{	rah2016GOSPA} is summarized in Subsection \ref{SS:Summary_GOSPA}. The GOSPA metric is an unnormalized version of OSPA with some scaling of the cardinality parameter. 
GOSPA actually helps to overcome the limitation of the  OSPA metric when one of the two sets is empty. However, the lack of normalization causes the GOSPA metric to increase when the sizes of the sets increase.  
These limitations are discussed in Subsection \ref{SS:IssueGOSPA}.
\subsection{Summary}\label{SS:Summary_GOSPA}
\begin{definition}[GOSPA metric]\label{D:GOSPA}
	Let $X$ and $Y$ be two finite sets. 
	For $p\geq 1$, $c>0,$ $0<\alpha\le 2$, the GOSPA metric,  $\bar{d}^{(c,\alpha)}_p(X,Y)=\bar{d}^{(c,\alpha)}_p(Y,X)$, is defined as follows for $|X|\le|Y|$, 
	\begin{eqnarray}\label{E:GOSPA}
		\left(\displaystyle\min_{\pi\in\Pi(X,Y)}\sum_{x\in X}\!\bar{d}^{(c)}
		(x,\pi(x))^p+\frac{c^p}{\alpha }(|Y|-|X|)\right)^{\frac{1}{p}}.&
	\end{eqnarray}
\end{definition}

\begin{Proposition}\label{P:PropGOSPA} Let $X$ and $Y$ be two finite sets. 
	For $p\geq 1$, $c>0$, $\alpha=2$, the GOSPA metric,  $\bar{d}^{(c,2)}_p(X,Y)=\bar{d}^{(c,2)}_p(Y,X)$, can be expressed as an optimization over assignment sets as for $|X|\le|Y|$,
	\begin{align}\label{E:PropGOSPA}		
		&\Bigg(\displaystyle\min_{\pi\in\Pi(X,Y)}\Bigg[\sum_{x\in X}\bar{\delta}\big(\bar{d}^{(c)}
		(x,\pi(x)),c\big)\bar{d}^{(c)}
		(x,\pi(x))^p\nonumber\\&+\frac{c^p}{2}\bigg(|X|+|Y|-2\sum_{x\in X}\bar{\delta}\big(\bar{d}^{(c)}
		(x,\pi(x)),c\big)\bigg)\Bigg]\Bigg)^{\frac{1}{p}}
	\end{align}
	where $\delta(k,h)$ is a Kronecker delta (i.e. $\delta(k,h)$ is $1$ if $k=h$ or $0$ if $k\ne h$) and $\bar{\delta}(k,h)=1-\delta(k,h)$.
\end{Proposition}
\subsection{Discussion of the GOSPA metric}\label{SS:IssueGOSPA}
The GOSPA metric overcomes 
 the limitations of the OSPA metric by removing the normalization of the OSPA metric and multiplying the second term of \eqref{E:OSPA} with $1/\alpha$ where $0<\alpha\le 2$. However, these alterations to the OSPA metric have an unexpected side-effect: GOSPA distance between two finite sets in general increases if one or both of the two sets increases in cardinality (details in Section \ref{S:Ana-COSPA}). 
 Furthermore, the cut-off $c$, being larger than the penalty given to each extra vector  $c/\sqrt[p]{\alpha}$ if $\alpha\ge 1$, makes the GOSPA metric even less useful (see Figure \ref{Fig:Ex_Empty_mEle}).  Definition \ref{D:GOSPA} shows that it is obviously that $\bar{d}^{(c,2)}_p(X,Y)\le \bar{d}^{(c,\alpha)}_p(X,Y)$ for all
 $0<\alpha\le 2$. Hence, it is not clear what an optimization over assignments sets mean in the Proposition \ref{P:PropGOSPA} but incidentally, Proposition \ref{P:PropGOSPA} is just a way to group the mismatched targets as the second term in \eqref{E:PropGOSPA} when $\alpha=2$ (the mismatched targets are two targets from two different sets that are further than or equal to $c$). If $\alpha\ne 2$, we cannot rewrite \eqref{E:GOSPA} in the form of \eqref{E:PropGOSPA}.

In order to overcome the limitations of the OSPA metric discussed in Section \ref{SS:LimitOSPA}, we propose a new metric, namely Complete OSPA (COSPA), in the next section. This metric not only overcomes the above limitations but also retains the advantages of the OSPA metric for evaluating cases where the OSPA metric gives reliable solutions.
\section{Complete Optimal Subpattern Assignment (COSPA) Metric}\label{S:COSPA_metric}
The Optimal Subpattern Assignment (OSPA) metric is very popular in multi-target filtering because it can evaluate the localization error and cardinality error between two sets of vectors. However, as discussed in Section \ref{SS:LimitOSPA} 
the OSPA metric between an empty set and a non-empty set is the same regardless of how many elements the non-empty set has. It is also indistinguishable between the cut-off distance and the cardinality penalty 
and discussed in Section \ref{SS:LimitOSPA}. Hence, in this section, we will derive a new metric which overcomes these shortcomings and still retains all the beneficial properties of the OSPA metric. The new metric, which is called Complete OSPA (COSPA), cuts off the distance between two vectors at 
$c$ ($c>0$), penalizes each extra of points in a bigger set by $\dot{c}$ ($\dot{c}\ge c$) and penalized the empty set error by $\xi$ ($0\le\xi\le 1$). In the rest of this section, we show that COSPA will overcome the shortcomings of the current metrics OSPA and COSPA.
\begin{definition}\label{D:COSPA}
	Let $X$ and $Y$ be two finite sets. For order parameter $p$ ($1\le p\le\infty$), cut-off parameter $c$ ($0<c$) and cardinality penalty parameter $\dot{c}$ ($\dot{c}\ge c$)  and empty-set parameter $0\le\xi\le 1$, the COSPA metric $d^{(\dot{c},\xi)}_{c,p}(Y,X)=d^{(\dot{c},\xi)}_{c,p}(X,Y)$ is defined as follows.
	\begin{itemize}
		\item If $X=Y=\emptyset$: $d^{(\dot{c},\xi)}_{c,p}(X,Y)=0.$
		\item Otherwise, for $p<\infty$, $d^{(\dot{c},\xi)}_{c,p}(X,Y)$ is for $|X|\le|Y|$
		\begin{subequations}\label{E:COSPA}
			\begin{eqnarray}
				&\hspace{-.5cm}	
				\Bigg(\displaystyle
				\dfrac{1}{|Y|}\bigg[\min_{\pi\in\Pi(X,Y)}\sum_{x\in X}\Big(c^p\delta\big(\bar{d}^{(c)}
				(x,\pi(x)),c\big)\label{E:DiffCOSPA2}\\
				&+\bar{\delta}\big(\bar{d}^{(c)}
				(x,\pi(x)),c\big)\bar{d}^{(c)}
				(x,\pi(x))^p\Big)\bigg]
				\label{E:DiffCOSPA1}\\
				&+\dot{c}^p\dfrac{|Y|-|X|}{|Y|}\label{E:DiffCOSPA3}\\
				& +\xi\dfrac{\delta(X,\emptyset)\bar{\delta}(Y,\emptyset)}{|Y|}\displaystyle \dot{c}^p(|Y|-1)\Bigg)^{\frac{1}{p}}\label{E:COSPA2}
			\end{eqnarray}
		\end{subequations}	
		\item if $p=\infty$, $d^{(\dot{c},\xi)}_{c,\infty}(X,Y)$ is 
		\begin{align}\label{E:COSPA_infty}
			\left\{
			\begin{array}{ll}
				\displaystyle\min_{\pi\in\Pi(X,Y)}\max_{x\in X}\bar{d}^{(c)}
				(x,\pi(x)) &\!\!\! \hbox{if $|X|=|Y|$} \\
				\dot{c}&  \hbox{if $|X|\ne |Y|$}
			\end{array}
			\right.
		\end{align}
	\end{itemize}
	The function $d^{(\dot{c},\xi)}_{c,p}(\cdot,\cdot)$ is called the COSPA metric of order $p$ with cut-off $c$, cardinality penalty $\dot{c}$ and empty-set error $\xi$.
\end{definition}
COSPA is proved as a metric in \cite{Tuyet2020} (note that $\dot{c}$ and $c$ are swapped in \cite{Tuyet2020}). If we choose $\dot{c}=c$ and $\xi=0$, COSPA is exactly OSPA. The term $\dot{c}$, which is the penalty for each cardinality error and differs from the cut off distance $c$, exists to overcome the limitation of OSPA for scenarios that distinguish between the cut-off distance from a cardinality error and are similar to the example shown in Figure \ref{Fig:ExSameExtraPoint}. 
In this example the OSPA distance between the two sets in Figure \eqref{Fig:ExSameExtraPoint1} is the same as the OSPA distance between the two sets in Figure \eqref{Fig:ExSameExtraPoint2}. In general, the OSPA distances for two pairs of finite sets are the same when the cardinality of a set of the first pair is smaller than the other set of that pair and the second pair is the same as the first pair with the smaller set now having an extra element whose distance to elements in the other set of that pair is larger than $c$. 

Note that the cut-off parameter $c$ is smaller than cardinality penalty $\dot{c}$ if the outline distance between two vectors is penalized less than each mismatched number of elements between two sets of vectors. If $\dot{c}=c$, Definition \ref{D:COSPA} (COSPA) is only different from Definition \ref{D:OSPA} (OSPA) \cite{Schuhmacher2008} by the term in \eqref{E:COSPA2}. The term in \eqref{E:COSPA2}  only exists to take into account the case when one of the two sets is empty. When both finite sets are non-empty the term \eqref{E:COSPA2} is zero, so COSPA is exactly the same as OSPA \cite{Schuhmacher2008} if $\dot{c}=c$.

If we do not wish to distinguish what vectors in $X$ are very far from their images in $Y$, 
$c^p\delta\big(\bar{d}^{(c)}
(x,\pi(x)),c\big)+\bar{\delta}\big(\bar{d}^{(c)}
(x,\pi(x)),c\big)\bar{d}^{(c)}
(x,\pi(x))^p=\bar{d}^{(c)}
(x,\pi(x))^p$ for 
$\pi\in\Pi(X,Y)$. 
Alternatively, $d^{(\dot{c},\xi)}_{c,p}(X,Y)$ in \eqref{E:COSPA} can be written for simplicity as follows if $|X|\le|Y|$ and $Y\ne\emptyset$.
\begin{subequations}\label{E:AltCOSPA}
	\begin{eqnarray}
		&\bigg[\displaystyle\frac{1}{|Y|}\bigg(\min_{\pi\in\Pi(X,Y)}\!\sum_{x\in X}\!\bar{d}^{(c)}
		\!(x,\pi(x))^p+\dot{c}^p(|Y|-|X|)\!\bigg)\label{E:AltCOSPA1}\\
		& 
		+\xi\dfrac{\delta(X,\emptyset)\bar{\delta}(Y,\emptyset)}{|Y|}\displaystyle \dot{c}^p(|Y|-1)\bigg]^{\frac{1}{p}}\label{E:AltCOSPA2}
	\end{eqnarray}
\end{subequations}	
\begin{rmk}\label{R:COSPA} From Definition \ref{D:OSPA}, Definition \ref{D:COSPA} and \eqref{E:AltCOSPA}, the following are true
	\begin{enumerate}
		\item If $X\ne\emptyset$ and $Y\ne\emptyset$, \eqref{E:COSPA} shows that the COSPA distance between two non-empty finite sets is smaller than or equal to $\dot{c}$. In this case, the COSPA distance \eqref{E:AltCOSPA} has the same forms as the OSPA distance \eqref{E:OSPA} because the term in \eqref{E:AltCOSPA2} does not exist. Furthermore, if $\dot{c}=c$, then the COSPA distance is the same as the OSPA distance, i.e. $d^{(\dot{c},\xi)}_{c,p}(X,Y)=\bar{d}^{(c)}_p(X,Y).$
		\item If $X=\emptyset$ or $Y=\emptyset$ but not both,
		\begin{align*}
			d^{(\dot{c},\xi)}_{c,p}(X,Y)=\dot{c}\left(1+\xi-\frac{\xi}{\max(|X|,|Y|)}\right)^\frac{1}{p}
		\end{align*}	
	\end{enumerate}
\end{rmk}
By \eqref{E:COSPA}, 
the COSPA assignment is the OSPA assignment. The COSPA assignment between $X$ and $Y$, $\pi^*$, is
\begin{align}\label{E:COSPA_Ass}
	\begin{array}{ll}
		\displaystyle\arg\min_{\pi\in\Pi(X,Y)}\sum_{x\in X}\bar{d}^{(c)}
		\!(x,\pi(x))^p
		,\!\!\!&\hbox{if  $|X|\le|Y|$;} \\
		\displaystyle\arg\min_{\pi\in\Pi(X,Y)}\sum_{y\in Y}\bar{d}^{(c)}
		\!(y,\pi(y))^p,&\hbox{otherwise.}
	\end{array}
\end{align}
\begin{rmk}\label{R:DefCOPSA} Assume that $X$ is the set of truth targets\footnote{In Remark \ref{R:DefCOPSA}, the term `target' and 'vector' are used interchangeably.} and $Y$ is the set of estimated targets. Definition \ref{D:COSPA} can be interpreted as follows.
	\begin{enumerate}
		\item The term 
		in \eqref{E:DiffCOSPA1} is actually the sum of distances between vectors in $X$ and their images in $Y$ by the one to one function $\pi^*$ if each of these distances is smaller than $c$. Each pair $(x,\pi^*(x))\in X\times Y$ is a pair of correctly associated targets (vectors) in $X$ and $Y$ if their distance is smaller than $c$. Note that $\pi^*\in\Pi(X,Y)$ is defined in \eqref{E:COSPA_Ass}. 
		\item The term in \eqref{E:DiffCOSPA2} is actually the sum of $|\gamma|$ distances between $|\gamma|$ vectors in $X$ and the $|\gamma|$  vectors in $Y$ by the one to one function $\pi^*$ when each of these distances is larger than or equal to $c$ where 
		\begin{align*}
			\gamma=\{(x,\pi^*(x))\in X\times Y:\bar{d}^{(c)}
			(x,\pi^*(x))=c\}.
		\end{align*}
		Each point in the map $\gamma$ is a pair of incorrectly associated targets if their distance is equal to or larger than $c$. 
		Alternatively, the target in $X$ 
		is called a missing target and the 
		corresponding estimated target in $Y$ via $\pi^*$
		is called a false target.
		
		\item The term in \eqref{E:DiffCOSPA3} is actually the cardinality error between $X$ and $Y$. If $|Y|>|X|$, the targets in $Y$ which do not associate with any targets in $X$ via $\pi^*$ are called false targets. 
		\item The term in \eqref{E:COSPA2} only exists if the smaller set is empty and the bigger set is not empty.		
	\end{enumerate}
\end{rmk}

The COSPA distance comprises of 3 components: COSPA localization error, COSPA outline error (i.e. the sum of all distances that are larger than or equal to $c$) and COSPA cardinality error. Here, the OSPA localization error is the sum of the COSPA localization error and the COSPA outline error. The COSPA cardinality error is the same as the OSPA cardinality error if both sets are not empty. If either $X$ or $Y$ is empty, the COSPA cardinality error is the COSPA distance (Remark \ref{R:COSPA}.2) while the OSPA cardinality error is the OSPA distance $c$. Hence, similar to the OSPA metric, the COSPA metric is interpreted as a $p-$th order per-target error. 
By \eqref{E:COSPA} for $1\le p<\infty$, COSPA localization \eqref{E:COSPALoc}, COSPA outline \eqref{E:COSPAOut} and cardinality errors \eqref{E:COSPACard} are given respectively as follows for $|X|\le|Y|$

\begin{subequations}
	\begin{align}
		&\bar{e}^{(c)}_{p,\text{loc}}(X,Y)=\bar{e}^{(c)}_{p,\text{loc}}(Y,X)\nonumber\\=&
		\bigg(\frac{1}{|Y|}\sum_{x\in X}\bar{\delta}\big(\bar{d}^{(c)}
		(x,\pi^*(x)),c\big)\bar{d}^{(c)}
		(x,\pi^*(x))^p\bigg)^{\frac{1}{p}},\label{E:COSPALoc}\\
		&\bar{e}^{(c)}_{p,\text{out}}(X,Y)=\bar{e}^{(c)}_{p,\text{out}}(Y,X)\nonumber\\=&c\left(\frac{\sum_{x\in X}\delta\big(\bar{d}^{(c)}
			(x,\pi^*(x)),c\big)}{|Y|}\right)^{\frac{1}{p}}\!\!\!\!,\label{E:COSPAOut}\\
		&\bar{e}^{(\xi)}_{\dot{c},p,\text{card}}(X,Y)=\bar{e}^{(\xi)}_{\dot{c},p,\text{card}}(Y,X)\nonumber\\&=\dot{c}\left(\frac{|Y|-|X|}{|Y|}+\xi\delta(X,\emptyset)\bar{\delta}(Y,\emptyset)\frac{|Y|-1}{|Y|}\right)^{\frac{1}{p}}\!\!\!\!\label{E:COSPACard}
	\end{align}
\end{subequations}
\section{Analysis of COSPA, GOSPA and OSPA metrics}\label{S:Ana-COSPA}
In  this section, we show the solution the COSPA metric offers to the limitations of OSPA and GOSPA mentioned in Sections \ref{SS:LimitOSPA}  and \ref{S:GOSPA} via simple scenarios. Furthermore, we also analyze the scenarios by comparing these solutions. Without loss of generality $\xi$ in Definition \ref{D:COSPA} is chosen as $1$ (i.e. $\xi=1$) and the order parameter $p=1$ for the three metrics.
\subsection{Effect of Cardinality Zero}\label{SS:CardZero}
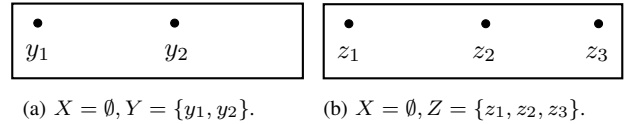
\begin{figure}[htbp!]
	\hspace{-.7cm}%
	\centering
	\subfloat[$X=\emptyset,Y=\{y_1,y_2\}$.]
	{\label{Fig:Ex_CutoffDifPenCarAlphaSm1_1}
		\centering{
			\scalebox{1}
			{\centering
				\pgfdeclarelayer{background} \pgfdeclarelayer{foreground}
\pgfsetlayers{background,main,foreground}
\par
\par
\tikzset{cross/.style={cross out, draw=black, minimum size=2*(#1-\pgflinewidth), inner sep=0pt, outer sep=0pt},
cross/.default={3pt}}
\begin{tikzpicture}[]

\draw (0,0.2) node[circle,draw=black, fill=black, inner sep=0pt,minimum size=3pt] (x1){};

\node (lx1)at (0,-.2){$y_1$};
\node[circle,draw=black, fill=black, inner sep=0pt,minimum size=3pt](x2)[right =1.7cm of x1] {};
\node(x3)[right =3.2cm of x1] {};
\node (lx3)at (3.35,-.2){};
\node (lx2)at (1.85,-.2){$y_2$};
%
%

\path (lx3.east |-x1.north)+(.05,+.2) node(cor1){};
  \path (lx1.west|-lx1.south)+(-.05,-.1) node(cor2){};
   \path[
    draw=black!100,thick]
            (cor1) rectangle  (cor2);
\end{tikzpicture}
		}}
	}\hspace{-1cm}\,
	\subfloat[$X=\emptyset,Z=\{z_1,z_2,z_3\}$.]
	{\label{Fig:Ex_CutoffDifPenCarAlphaSm1_2}
		\centering{
			\scalebox{1}
			{\centering
				\pgfdeclarelayer{background} \pgfdeclarelayer{foreground}
\pgfsetlayers{background,main,foreground}
\par
\par
\tikzset{cross/.style={cross out, draw=black, minimum size=2*(#1-\pgflinewidth), inner sep=0pt, outer sep=0pt},
cross/.default={3pt}}
\begin{tikzpicture}[]

\draw (0,0.2) node[circle,draw=black, fill=black, inner sep=0pt,minimum size=3pt] (x1){};

\node (lx1)at (0,-.2){$z_1$};
\node[circle,draw=black, fill=black, inner sep=0pt,minimum size=3pt](x2)[right =1.7cm of x1] {};
\node[circle,draw=black, fill=black, inner sep=0pt,minimum size=3pt](x3)[right =3.2cm of x1] {};
\node (lx3)at (3.3,-.2){$z_3$};
\node (lx2)at (1.8,-.2){$z_2$};

%
%
\path (lx3.east |-x1.north)+(.05,+.2) node(cor1){};
  \path (lx1.west|-lx1.south)+(-.05,-.1) node(cor2){};
   \path[
    draw=black!100,thick]
            (cor1) rectangle  (cor2);
\end{tikzpicture}
	}}}
	\caption{The OSPA metric $\bar{d}^{(c)}_p(\emptyset,Y)=\frac{2c}{2}=c= \frac{3c}{3}=\bar{d}^{(c)}_p(\emptyset,Z)$. OSPA assigns the same value to the distance between any non-empty set and an empty set.}
	\label{Fig:Ex_CutoffDifPenCarAlphaSm1}
\end{figure}

\begin{figure}[htbp!]
	\hspace{-.7cm}%
	\centering\subfloat[$Z=\emptyset$]
	{\label{Fig:Ex_Empty_mEle_1}
		\centering{
			\scalebox{.9}
			{\centering
				\pgfdeclarelayer{background} \pgfdeclarelayer{foreground}
\pgfsetlayers{background,main,foreground}
\par
\par
\tikzset{cross/.style={cross out, draw=black, minimum size=2*(#1-\pgflinewidth), inner sep=0pt, outer sep=0pt},
	cross/.default={3pt}}
\begin{tikzpicture}[]

\draw (0,0) node[circle,draw=black, fill=black, inner sep=0pt,minimum size=2.5pt](x1){};
\draw (0,-.3) node {$x_1$};

\draw (0.8,0) node[circle,draw=black, fill=black, inner sep=0pt,minimum size=2.5pt](x2) {};
\draw (0.8,-.3)node {$x_2$};

\draw (1.6,0) node[circle,draw=black, fill=black, inner sep=0pt,minimum size=2.5pt](xi) {};
\draw (1.6,-.3) node {$x_i$};

\draw (2.4,0) node[circle,draw=black, fill=black, inner sep=0pt,minimum size=2.5pt](xn-1) {};

\draw (3.2,0) node[circle,draw=black, fill=black, inner sep=0pt,minimum size=2.5pt](xn) {};
\draw (3.3,-.3) node {$x_m$};

\path (xn.east |-x1.north)+(.35,+.25) node(cor1){};
\path (x1.west|-x1.south)+(-.35,-.5) node(cor2){};
\path[
draw=black!100,thick]
(cor1) rectangle  (cor2);
\end{tikzpicture}
	}}}\hspace{-.3cm}\,
	\subfloat[$Y=\{y_1,\ldots,y_m\}$]
	{\label{Fig:Ex_Empty_mEle_2}
		\centering{
			\scalebox{.9}
			{\centering
				\pgfdeclarelayer{background} \pgfdeclarelayer{foreground}
\pgfsetlayers{background,main,foreground}
\par
\par
\tikzset{cross/.style={cross out, draw=black, minimum size=2*(#1-\pgflinewidth), inner sep=0pt, outer sep=0pt},
	cross/.default={3pt}}
\begin{tikzpicture}[]

\draw (0,0) node[circle,draw=black, fill=black, inner sep=0pt,minimum size=2.5pt](x1){};
\draw (0,-.3) node {$x_1$};

\draw (0.8,0) node[circle,draw=black, fill=black, inner sep=0pt,minimum size=2.5pt](x2) {};
\draw (0.8,-.3)node {$x_2$};

\draw (1.6,0) node[circle,draw=black, fill=black, inner sep=0pt,minimum size=2.5pt](xi) {};
\draw (1.6,-.3) node {$x_i$};

\draw (2.4,0) node[circle,draw=black, fill=black, inner sep=0pt,minimum size=2.5pt](xn-1) {};

\draw (3.2,0) node[circle,draw=black, fill=black, inner sep=0pt,minimum size=2.5pt](xn) {};
\draw (3.3,-.3) node {$x_m$};

\draw (0,1) node[circle,draw=black, fill=black, inner sep=0pt,minimum size=2.5pt](y1){};
\draw (0,1.2) node {$y_1$};

\draw (.8,1) node[circle,draw=black, fill=black, inner sep=0pt,minimum size=2.5pt](y2){};
\draw (.8,1.2) node {$y_2$};

\draw (1.6,1) node[circle,draw=black, fill=black, inner sep=0pt,minimum size=2.5pt](yi){};
\draw (1.6,1.2) node {$y_i$};

\draw (2.4,1) node[circle,draw=black, fill=black, inner sep=0pt,minimum size=2.5pt](yn-1) {};

\draw (3.2,1) node[circle,draw=black, fill=black, inner sep=0pt,minimum size=2.5pt](yn){};
\draw (3.3,1.2) node {$y_m$};
%
%
%
%

\node(temdis11)  [above =.33cm of x1] {};
\node (dis2) [right =-.15cm of temdis11]{\scriptsize$\eta$};
\draw[-,thin,black,dotted](x1)edge node[swap,near start]{}  (y1);


\path (yn.east |-yn.north)+(.5,+.45) node(cor1){};
\path (x1.west|-x1.south)+(-.35,-.5) node(cor2){};
\path[
draw=black!100,thick]
(cor1) rectangle  (cor2);
\end{tikzpicture}
	}}}
	\caption{The ground truth is $X=\{x_1,\ldots,x_m\}$, $m>1$. In Figure \ref{Fig:Ex_Empty_mEle_1}, an algorithm estimates no target, i.e. $Z=\emptyset$, while an algorithm in Figure \ref{Fig:Ex_Empty_mEle_2} estimates $m$ targets, i.e $Y=\{y_1,y_2,\ldots,y_{m}\}$.}
	\label{Fig:Ex_Empty_mEle}
\end{figure}
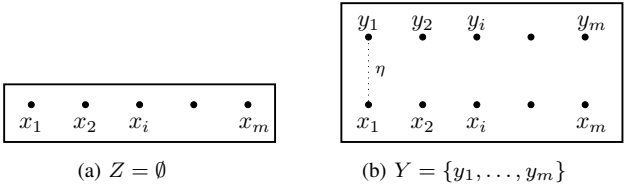

As discussed in Section \ref{SS:LimitOSPA}, the OSPA metric gives the same result when one of the two arguments is empty. In Figure \ref{Fig:Ex_CutoffDifPenCarAlphaSm1} the OSPA distance between the non-empty set $Y$ and $\emptyset$ 
is the same as the OSPA distance between the non-empty set $Z$ and $\emptyset$. The GOSPA and COSPA metrics give smaller values for  $(\emptyset, Y)$ than $(\emptyset, Z)$. As a result, a set with two vectors $Y$ is closer to $\emptyset$ than a set with three vectors $Z$. This is a natural (meaningful) physical interpretation. Detailed explanations are given in Table \ref{table:EmptyNonEmpty}.
\renewcommand{\arraystretch}{1.7}
\begin{table}[htpb!]%
	\caption{\textbf{Analysis of the three metrics for Figure \ref{Fig:Ex_CutoffDifPenCarAlphaSm1}}}
	\begin{center}
		\begin{tabular}
			{|l||l|l|l|}\hline
			\!Metric\!&\!Fig.\ref{Fig:Ex_CutoffDifPenCarAlphaSm1_1} ($Y$)&Fig. \ref{Fig:Ex_CutoffDifPenCarAlphaSm1_2} ($Z$)&Which one is closer to $\emptyset$? \\ \hline\hline
			\!OSPA &\!$c$ & \!$c$& They are the same \\ \hline
			\!GOSPA&\!$c\frac{2}{\alpha}$\!& \!$c\frac{3}{\alpha}$&$Y$\\\hline
			\!COSPA\!&\!$\dot{c}\frac{3}{2}$ &\!$\dot{c}\frac{5}{3}$&$Y$\\ \hline
			Intuition\footnote{`Intuition' usually corresponds to a surveillance operator's likely needs, i.e. the cardinality penalty is larger than or equal to the distance cut off due to $\dot{c}\ge c$.}& & & $Y$\\ \hline
		\end{tabular}
		\label{table:EmptyNonEmpty}
	\end{center}
\end{table}
The COSPA metric is different from the other two metrics when comparing one algorithm producing an empty set and another algorithm giving a non-empty set as an output. Compared with the non-empty set ground truth, COSPA gives an algorithm with the empty set a bigger value than an algorithm with the non-empty set. The distance between two empty sets is zero. It is clear that the three metrics will give the algorithm with the empty set a smaller distance when the ground truth is empty. Figure \ref{Fig:Ex_Empty_mEle} and Table \ref{table:Fig:Ex_Empty_mEle} give an example when the ground truth is not empty. GOSPA  concludes that the empty set $Z$ is the closer estimate to $X$ than the non empty set $Y$ if $\frac{c}{\alpha}\le\eta$ and $\alpha>1$. These results depend on the choice of cut-off $c$. If $c\le\eta$, OSPA concludes that $Y$ is as good as $Z$ for estimating $X$, otherwise $Y$ is the better estimate of $X$ than $Z$. The COSPA metric concludes that $Y$ is the better estimate of $X$ than $Z$ no matter what values are chosen for the cut-off $c$ and cardinality  error penalty $\dot{c}$ (see more detail in Table \ref{table:Fig:Ex_Empty_mEle}). This is a 
natural (meaningful) physical interpretation.

\renewcommand{\arraystretch}{1.7}
\begin{center}
	\begin{table}[htpb!]%
		\caption{\textbf{Analysis of the three metrics 
				for Figure \ref{Fig:Ex_Empty_mEle}}}
		\begin{center}
			\begin{tabular}
				{|l||l|l|l|}\hline
				Metric\!&Fig. \ref{Fig:Ex_Empty_mEle_1}($Z$)&Fig. \ref{Fig:Ex_Empty_mEle_2} ($Y$)&Which one 
				is closer to $X$? \\ \hline\hline
				OSPA &$c$ & $\min(c,\eta)$& 
				$Y$ if $\boldsymbol{\eta<c}$
				\\ \hline
				GOSPA&$c\frac{m}{\alpha}$& $m\min(c,\eta)$&
				$Y$  if $\boldsymbol{\eta<\frac{c}{\alpha}}$
				\\ \hline
				COSPA&$\dot{c}\left(2-\frac{1}{m}\right)$ & $\min(c,\eta)
				$&$Y$\\ \hline
				Intuition& & &$Y$\\ \hline
			\end{tabular}
			\label{table:Fig:Ex_Empty_mEle}
		\end{center}
	\end{table}
\end{center}
\renewcommand{\arraystretch}{1}
\subsection{Effect of Choice of Cut-off and Cardinality  Penalty }\label{SS:Cut-off_PenalErr}
\begin{figure}[htbp!]
	\hspace{-.7cm}%
	\centering
	\subfloat[$Y=\{y_1,y_2\}$.]
	{\label{Fig:ExSameExtraPoint1}
		\centering{
			\scalebox{1.25}
			{\centering
				\pgfdeclarelayer{background} \pgfdeclarelayer{foreground}
\pgfsetlayers{background,main,foreground}
\par
\par
\tikzset{cross/.style={cross out, draw=black, minimum size=2*(#1-\pgflinewidth), inner sep=0pt, outer sep=0pt},
	cross/.default={3pt}}
\begin{tikzpicture}[]

\draw (0,0) node[circle,draw=black, fill=black, inner sep=0pt,minimum size=2.5pt](x1){};
\draw (0,-.3) node {$x_1$};

\draw (0.6,0) node[circle,draw=black, fill=black, inner sep=0pt,minimum size=2.5pt](x2) {};
\draw (0.6,-.3)node {$x_2$};

\draw (1.2,0) node[circle,draw=black, fill=black, inner sep=0pt,minimum size=2.5pt](x3) {};
\draw (1.2,-.3) node {$x_3$};

\draw (0,.4) node[circle,draw=black, fill=black, inner sep=0pt,minimum size=2.5pt](y1){};
\draw (0,.6) node {$y_1$};

\draw (.6,.4) node[circle,draw=black, fill=black, inner sep=0pt,minimum size=2.5pt](y2){};
\draw (.6,.6) node {$y_2$};


\node(temdis11)  [above =.05cm of x1] {};
\node (dis2) [right =-.15cm of temdis11]{\scriptsize$\Delta$};
\draw[-,thin,black,dashed](x1)edge node[swap,near start]{}  (y1);
\draw[-,thin,black,dashed](x2)edge node[swap,near start]{}  (y2);

\path (x3.east |-y2.north)+(.5,+.45) node(cor1){};
\path (x1.west|-x1.south)+(-.35,-.5) node(cor2){};
\path[
draw=black!100,thick]
(cor1) rectangle  (cor2);
\end{tikzpicture}
		}}
	}\hspace{-.8cm}\,
	\subfloat[$Z=\{y_1,y_2,z\}$.]
	{\label{Fig:ExSameExtraPoint2}
		\centering{
			\scalebox{1.25}
			{\centering
				\pgfdeclarelayer{background} \pgfdeclarelayer{foreground}
\pgfsetlayers{background,main,foreground}
\par
\par
\tikzset{cross/.style={cross out, draw=black, minimum size=2*(#1-\pgflinewidth), inner sep=0pt, outer sep=0pt},
	cross/.default={3pt}}
\begin{tikzpicture}[]

\draw (0,0) node[circle,draw=black, fill=black, inner sep=0pt,minimum size=2.5pt](x1){};
\draw (0,-.3) node {$x_1$};

\draw (0.6,0) node[circle,draw=black, fill=black, inner sep=0pt,minimum size=2.5pt](x2) {};
\draw (0.6,-.3)node {$x_2$};

\draw (1.2,0) node[circle,draw=black, fill=black, inner sep=0pt,minimum size=2.5pt](x3) {};
\draw (1.2,-.3) node {$x_3$};

\draw (0,.4) node[circle,draw=black, fill=black, inner sep=0pt,minimum size=2.5pt](y1){};
\draw (0,.6) node {$y_1$};

\draw (.6,.4) node[circle,draw=black, fill=black, inner sep=0pt,minimum size=2.5pt](y2){};
\draw (.6,.6) node {$y_2$};

\draw (2.8,.4) node[circle,draw=black, fill=black, inner sep=0pt,minimum size=2.5pt](y2n1){};
\draw (3,.6) node {$z$};

\node(temdis11)  [above =.05cm of x1] {};
\node (dis2) [right =-.15cm of temdis11]{\scriptsize$\Delta$};
\draw[-,thin,black,dashed](x1)edge node[swap,near start]{}  (y1);
\draw[-,thin,black,dashed](x2)edge node[swap,near start]{}  (y2);
\draw[-,thin,black,dashed](x3)edge node[swap,near start]{}  (y2n1);

\path (y2n1.east |-y2n1.north)+(.5,+.45) node(cor1){};
\path (x1.west|-x1.south)+(-.35,-.5) node(cor2){};
\path[
draw=black!100,thick]
(cor1) rectangle  (cor2);
\end{tikzpicture}
	}}}
	\caption{$X=\{x_1,x_2,x_3\}$ where $d(x_i,z)\ge c>\Delta$ for $i=1,2,3$. 
		The OSPA metric $\bar{d}^{(c)}_p(X,Y)=\frac{2\Delta+c}{3}= \bar{d}^{(c)}_p(X,Z)$. 
	OSPA gives the same value for the distance between $X$ and $Y$ and for the distance between $X$ and $Z$.}
	\label{Fig:ExSameExtraPoint}
\end{figure}
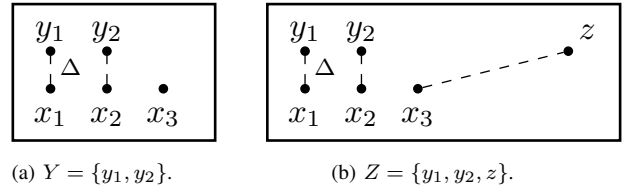

\begin{figure}[htbp!]
	\hspace{-.7cm}%
	\centering
	\subfloat[
	$Y_a=\emptyset$.]
	{\label{Fig:Ex_RemDe_0}
		\centering{
			\scalebox{.9}
			{\centering
				\pgfdeclarelayer{background} \pgfdeclarelayer{foreground}
\pgfsetlayers{background,main,foreground}
\par
\par
\tikzset{cross/.style={cross out, draw=black, minimum size=2*(#1-\pgflinewidth), inner sep=0pt, outer sep=0pt},
cross/.default={3pt}}
\begin{tikzpicture}[]

\draw (0,0.2) node[cross] (x1){};

\node (lx1)at (0,-.2){$x_1$};
\node(x3)[right =.7cm of x1] {};
\draw (1.1,.2) node[cross] (x2){};
\node[above =2cm of x3] (y3) {};
\node(y2) at (1.1,1)
{};
\node (ly3) at (1,2.7){};
\node (ly2) at (1.1,1.3){};
\node (lx2)at (1.1,-.2){$x_2$};

%
\path (ly2.east |-ly3.north)+(.14,+.12) node(cor1){};
  \path (lx1.west|-lx1.south)+(-.01,-.1) node(cor2){};
   \path[
    draw=black!100,thick]
            (cor1) rectangle  (cor2);
\end{tikzpicture}
	}}}\hspace{-.7cm}\,
	\subfloat[
	$Y_b=\{y_1\}$.]
	{\label{Fig:Ex_RemDe_1}
		\centering{
			\scalebox{.9}
			{\centering
				\pgfdeclarelayer{background} \pgfdeclarelayer{foreground}
\pgfsetlayers{background,main,foreground}
\par
\par
\tikzset{cross/.style={cross out, draw=black, minimum size=2*(#1-\pgflinewidth), inner sep=0pt, outer sep=0pt},
cross/.default={3pt}}
\begin{tikzpicture}[]

\draw (0,0.2) node[cross] (x1){};

\node[circle,draw=black, fill=black, inner sep=0pt,minimum size=3pt] (y1) at (0,1) {};
\node (ly1) at (0,1.3){$y_1$};
\node (lx1)at (0,-.2){$x_1$};
\node(x3)[right =.7cm of x1] {};
\draw (1.1,.2) node[cross] (x2){};
\node[above =2cm of x3] (y3) {};
\node(y2) at (1.1,1)
{};
\node (ly3) at (1,2.7){};
\node (ly2) at (1.1,1.3){};
\node (lx2)at (1.1,-.2){$x_2$};

\node(temdis11)  [below =.2cm of y1] {};
 \node (dis1) [right =.05cm of temdis11] {\scriptsize$\Delta$};
 \draw[-,thin,black,dashed](x1)edge node[swap,near start]{}  (y1);
%
\path (ly2.east |-ly3.north)+(.14,+.12) node(cor1){};
  \path (lx1.west|-lx1.south)+(-.01,-.1) node(cor2){};
   \path[
    draw=black!100,thick]
            (cor1) rectangle  (cor2);
\end{tikzpicture}
	}}}\hspace{-.6cm}\,
	\subfloat[
	$Y_c\!=\!\{\!y_1\!,y_2\!\}$.]
	{\label{Fig:Ex_RemDe_2}
		\centering{
			\scalebox{.9}
			{\centering
				\pgfdeclarelayer{background} \pgfdeclarelayer{foreground}
\pgfsetlayers{background,main,foreground}
\par
\par
\tikzset{cross/.style={cross out, draw=black, minimum size=2*(#1-\pgflinewidth), inner sep=0pt, outer sep=0pt},
cross/.default={3pt}}
\begin{tikzpicture}[]

\draw (0,0.2) node[cross] (x1){};

\node[circle,draw=black, fill=black, inner sep=0pt,minimum size=3pt] (y1) at (0,1) {};
\node (ly1) at (0,1.3){$y_1$};
\node (lx1)at (0,-.2){$x_1$};
\node(x3)[right =.7cm of x1] {};
\draw (1.1,.2) node[cross] (x2){};
\node[above =2cm of x3] (y3) {};
\node[circle,draw=black, fill=black, inner sep=0pt,minimum size=3pt] (y2) at (1.1,1)
{};
\node (ly3) at (1,2.7){};
\node (ly2) at (1.1,1.3){$y_2$};
\node (lx2)at (1.1,-.2){$x_2$};

\node(temdis11)  [below =.2cm of y1] {};
 \node (dis1) [right =.05cm of temdis11] {\scriptsize$\Delta$};
 \draw[-,thin,black,dashed](x1)edge node[swap,near start]{}  (y1);

 \node(temdis)  [below =.2cm of y2] {};
 \node (dis2) [right =.05cm of temdis]{};
 \draw[-,thin,black,dashed](x2)edge node[swap,near start]{}  (y2);
\path (ly2.east |-ly3.north)+(.14,+.12) node(cor1){};
  \path (lx1.west|-lx1.south)+(-.01,-.1) node(cor2){};
   \path[
    draw=black!100,thick]
            (cor1) rectangle  (cor2);
\end{tikzpicture}
	}}}\hspace{-.3cm}\,
	\subfloat[
	$Y_d\!=\!\{\!y_1\!,y_2\!,y_3\!\}$.]
	{\label{Fig:Ex_RemDe_3}
		\centering{
			\scalebox{.9}
			{\centering
				\pgfdeclarelayer{background} \pgfdeclarelayer{foreground}
\pgfsetlayers{background,main,foreground}
\par
\par
\tikzset{cross/.style={cross out, draw=black, minimum size=2*(#1-\pgflinewidth), inner sep=0pt, outer sep=0pt},
cross/.default={3pt}}
\begin{tikzpicture}[]

\draw (0,0.2) node[cross] (x1){};

\node[circle,draw=black, fill=black, inner sep=0pt,minimum size=3pt] (y1) at (0,1) {};
\node (ly1) at (0,1.3){$y_1$};
\node (lx1)at (0,-.2){$x_1$};
\node(x3)[right =.7cm of x1] {};
\draw (1.1,.2) node[cross] (x2){};
\node[circle,draw=black, fill=black, inner sep=0pt,minimum size=3pt,above =2cm of x3] (y3) {};
\node[circle,draw=black, fill=black, inner sep=0pt,minimum size=3pt] (y2) at (1.1,1)
{};
\node (ly3) at (1,2.7){$y_3$};
\node (ly2) at (1.1,1.3){$y_2$};
\node (lx2)at (1.1,-.2){$x_2$};

\node(temdis11)  [below =.2cm of y1] {};
 \node (dis1) [right =.05cm of temdis11] {\scriptsize$\Delta$};
 \draw[-,thin,black,dashed](x1)edge node[swap,near start]{}  (y1);

 \node(temdis)  [below =.2cm of y2] {};
 \node (dis2) [right =.05cm of temdis]{};
 \draw[-,thin,black,dashed](x2)edge node[swap,near start]{}  (y2);
\path (ly2.east |-ly3.north)+(.01,+.01) node(cor1){};
  \path (lx1.west|-lx1.south)+(-.01,-.1) node(cor2){};
   \path[
    draw=black!100,thick]
            (cor1) rectangle  (cor2);
\end{tikzpicture}
	}}}
	\caption{
		 $\Delta<c$, $X=\{x_1,x_2\}$ and the distances $d(x_1,y_1)=d(x_2,y_2)=\Delta$ and $d(x_i,y_3)>\Delta$ for $i=1,2$.}
	\label{Fig:Ex_RemDe}
\end{figure}
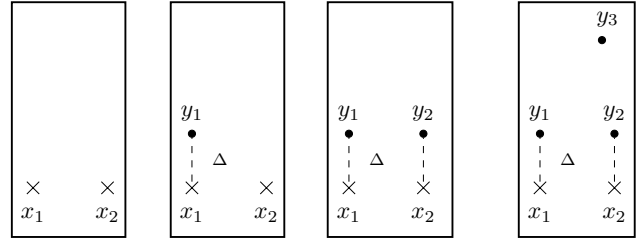

In the OSPA metric, the cut-off
 distance  $c$ between two vectors is the same as the penalty for each extra vector. For the COSPA metric, this cut-off distance $c$ 
  is smaller than or equal to the penalty for each extra vector, $\dot{c}$. This is not the case for the GOSPA metric \eqref{E:GOSPA} where the cut-off distance $c$ between two vectors is larger than or equal to the penalty for each extra vector, $c/\sqrt[p]{\alpha}$. 
This choice makes the GOSPA metric intuitively unreliable\footnote{If a metric is `unreliable', it occasionally assign a large value to the distance between two sets of vectors that are intuitively close.} for most other scenarios where both sets are not empty. 
GOSPA gives the same value for the two scenarios in Figure \ref{Fig:ExSameExtraPoint} if $\alpha = 1$; the smaller or bigger value for scenario \ref{Fig:ExSameExtraPoint1} than \ref{Fig:ExSameExtraPoint2} if $\alpha < 1$ or $\alpha > 1$ respectively. If $\alpha>1$, which is the preferred distance choice of the author, GOSPA gives the smaller, same or bigger distance for scenario \ref{Fig:Ex_RemDe_0} than \ref{Fig:Ex_RemDe_1} if $c> \Delta> c/\alpha$, $\Delta= c/\alpha$, or $\Delta<c/\alpha$ respectively.

As long as $\Delta<c$, OSPA gives the same value for the two scenarios in Figure \ref{Fig:ExSameExtraPoint} while COSPA gives a smaller value for scenario \ref{Fig:ExSameExtraPoint1} than \ref{Fig:ExSameExtraPoint2} if $\dot{c}>c$ and the same value if $\dot{c}=c$. This is summarized in Table \ref{table:ExSameExtraPoint}. Both OSPA and COSPA give the smaller value for scenario \ref{Fig:Ex_RemDe_1} than for \ref{Fig:Ex_RemDe_0}, agreeing with intuitive thinking. This is shown in Table \ref{table:Fig:Ex_RemDe}. Intuitively, $Y_c$ is the closest to $X$ as given by OSPA and COSPA. However, the GOSPA metric only gives the same conclusion if $\Delta<c/\alpha$. If $\alpha=2$, which the authors \cite{rah2016GOSPA} claim is the optimal solution for the GOSPA metric, GOSPA gives  $Y_a$ as the closest to $X$ if $\Delta\ge c/\alpha$, which contradicts intuitive thinking. The behaviour of these three metrics for the scenarios in Figure \ref{Fig:Ex_RemDe} are summarized in Table \ref{table:Fig:Ex_RemDe}.

\renewcommand{\arraystretch}{1.7}
\begin{table*}[htpb]%
	\caption{\textbf{Analysis of the three metrics 
		for	Figure \ref{Fig:ExSameExtraPoint}}}
	\begin{center}
		\begin{tabular}
			{|l||l|l|l|l|}\hline
			\!Metric\!&\!Fig.\ref{Fig:ExSameExtraPoint1} ($Y$)&\!Fig.\ref{Fig:ExSameExtraPoint2} ($Z$)&\!\!Which one ($Y$ or $Z$) is closer to $X$?&Explanation\\ \hline\hline
			\!OSPA &\!$\frac{2\Delta^p+c^p}{3}$ & \!$\frac{2\Delta+c}{3}$&They have the same distance to $X$ &cut-off $=$ cardinality penalty\\ \hline
			\!GOSPA&\!$2\Delta+\frac{1}{\alpha}c$\!& \!$2\Delta+c$&$Y$ if $\alpha>1$; $Z$ if $\alpha<1$ &cut-off $\ne$ cardinality penalty ($c\ne\frac{c}{\alpha}$)\\\hline
			\!COSPA\!&\!$\frac{2\Delta+\dot{c}}{3}$ &\!$\frac{2\Delta+c}{3}$&$Z$ if $c<\dot{c}$& cut-off $\le$ cardinality penalty ($c\le\dot{c}$)\\ \hline
		\end{tabular}
		\label{table:ExSameExtraPoint}
	\end{center}
\end{table*}
\renewcommand{\arraystretch}{1}

\renewcommand{\arraystretch}{1.7}
\begin{center}
	\begin{table*}[htpb!]%
		\caption{\textbf{Analysis of the three metrics and intuitive thinking to evaluate the sets shown in Figure \ref{Fig:Ex_RemDe}}}
		\begin{center}
			\begin{tabular}
				{|l||l|l|l|l|l|}\hline
				\!Metric\!&\!Fig.\ref{Fig:Ex_RemDe_0} ($Y_a$)&Fig.\ref{Fig:Ex_RemDe_1} ($Y_b$)&\!Fig. \ref{Fig:Ex_RemDe_2} ($Y_c$)&\!Fig.\ref{Fig:Ex_RemDe_3}  ($Y_d$)&Which one is the closest to $X$?\\ \hline\hline
				\!OSPA &\!$c$ & \!$\frac{c+\Delta}{2}$ &\!$\Delta$&\!$\frac{2\Delta+c}{3}$&$Y_c$\\ \hline
				\!GOSPA&\!$c\frac{2}{\alpha}$\!& \!$\Delta+\frac{c}{\alpha}$&\!$2\Delta$&\!$2\Delta+\frac{c}{\alpha}$&
				$Y_c$ if $\boldsymbol{\Delta< c\frac{1}{\alpha}}$ or $Y_a$ if $\boldsymbol{\Delta\ge c\frac{1}{\alpha}}$
				\\ \hline
				\!COSPA\!&\!$\dot{c}\frac{3}{2}$ & \!$\frac{\dot{c}+\Delta}{2}$ &\!$\Delta$&\!$\frac{2\Delta+\dot{c}}{3}$&$Y_c$\\ \hline
				Intuition& & &  &&$Y_c$\\ \hline
			\end{tabular}
			\label{table:Fig:Ex_RemDe}
		\end{center}
	\end{table*}
\end{center}
\renewcommand{\arraystretch}{1}

\subsection{Importance of Normalization}\label{SS:Normalization}
\begin{figure}[htbp!]
	\hspace{-.7cm}%
	\centering\subfloat[]
	{\label{Fig:Ex_Biggerset_0}
		\centering{
			\scalebox{.9}
			{\centering
				\pgfdeclarelayer{background} \pgfdeclarelayer{foreground}
\pgfsetlayers{background,main,foreground}
\par
\par
\tikzset{cross/.style={cross out, draw=black, minimum size=2*(#1-\pgflinewidth), inner sep=0pt, outer sep=0pt},
cross/.default={3pt}}
\begin{tikzpicture}[]

\draw (0,0) node[circle,draw=black, fill=black, inner sep=0pt,minimum size=2.5pt]
(x1){};
\draw (0,-.2) node {$l_1$};

\draw (0,2) node[circle,draw=black, fill=black, inner sep=0pt,minimum size=2.5pt](x2) {};

\draw (1.1,0) node[circle,draw=black, fill=black, inner sep=0pt,minimum size=2.5pt](y1){};
\draw (1.1,-.2) node {$l_2$};

\draw (1.1,2) node[circle,draw=black, fill=black, inner sep=0pt,minimum size=2.5pt](y2){};

\node(temdis11)  [above =.2cm of x1] {};
\node (dis2) [right =.3cm of temdis11] {};
\node (dis1) [below =-.1cm of dis2]{\scriptsize$\Delta$};
\draw[-,thin,black,dotted](x1)edge node[swap,near start]{}  (y1);

\draw[-,thick,black](x1)edge node[swap,near start]{}  (x2);
\draw[-,thick,black](y1)edge node[swap,near start]{}  (y2);

\path (y1.east |-x2.north)+(.25,+.35) node(cor1){};
\path (x1.west|-x1.south)+(-.25,-.35) node(cor2){};
\path[
draw=black!100,thick]
(cor1) rectangle  (cor2);

%
%
%
%
\end{tikzpicture}
	}}}\hspace{-.7cm}\,
	\subfloat[]
	{\label{Fig:Ex_Biggerset_1}
		\centering{
			\scalebox{.9}
			{\centering
				\pgfdeclarelayer{background} \pgfdeclarelayer{foreground}
\pgfsetlayers{background,main,foreground}
\par
\par
\tikzset{cross/.style={cross out, draw=black, minimum size=2*(#1-\pgflinewidth), inner sep=0pt, outer sep=0pt},
	cross/.default={3pt}}
\begin{tikzpicture}[]

\draw (0,0) node[circle,draw=black, fill=black, inner sep=0pt,minimum size=2.5pt](x1){};
\draw (0,-.2) node {$x_1$};

\draw (0,2) node[circle,draw=black, fill=black, inner sep=0pt,minimum size=2.5pt](x2) {};
\draw (0,2.15) node {$x_2$};

\draw (1.1,0) node[circle,draw=black, fill=black, inner sep=0pt,minimum size=2.5pt](y1){};
\draw (1.1,-.2) node {$y_1$};

\draw (1.1,2) node[circle,draw=black, fill=black, inner sep=0pt,minimum size=2.5pt](y2){};
\draw (1.1,2.15) node {$y_2$};

\node(temdis11)  [above =.2cm of x1] {};
\node (dis2) [right =.3cm of temdis11] {};
\node (dis1) [below =-.1cm of dis2]{\scriptsize$\Delta$};
\draw[-,thin,black,dotted](x1)edge node[swap,near start]{}  (y1);

\draw[-,thick,black,dashed](x1)edge node[swap,near start]{}  (x2);
\draw[-,thick,black,dashed](y1)edge node[swap,near start]{}  (y2);

\path (y2.east |-x2.north)+(.2,+.35) node(cor1){};
\path (x1.west|-x1.south)+(-.25,-.35) node(cor2){};
\path[
draw=black!100,thick]
(cor1) rectangle  (cor2);
\end{tikzpicture}
	}}}\hspace{-.6cm}\,
	\subfloat[]
	{\label{Fig:Ex_Biggerset_2}
		\centering{
			\scalebox{.9}
			{\centering
				\pgfdeclarelayer{background} \pgfdeclarelayer{foreground}
\pgfsetlayers{background,main,foreground}
\par
\par
\tikzset{cross/.style={cross out, draw=black, minimum size=2*(#1-\pgflinewidth), inner sep=0pt, outer sep=0pt},
	cross/.default={3pt}}
\begin{tikzpicture}[]

\draw (0,0) node[circle,draw=black, fill=black, inner sep=0pt,minimum size=2.5pt](x1){};
\draw (0,-.2) node {$x_1$};

\draw (0,1) node[circle,draw=black, fill=black, inner sep=0pt,minimum size=2.5pt](x2) {};
\draw (-0.25,1) node {$x_2$};

\draw (0,2) node[circle,draw=black, fill=black, inner sep=0pt,minimum size=2.5pt](x3) {};
\draw (0,2.15) node {$x_3$};

\draw (1.1,0) node[circle,draw=black, fill=black, inner sep=0pt,minimum size=2.5pt](y1){};
\draw (1.1,-.2) node {$y_1$};

\draw (1.1,1) node[circle,draw=black, fill=black, inner sep=0pt,minimum size=2.5pt](y2){};
\draw (.85,1) node {$y_2$};

\draw (1.1,2) node[circle,draw=black, fill=black, inner sep=0pt,minimum size=2.5pt](y3){};
\draw (1.1,2.15) node {$y_3$};

\node(temdis11)  [above =.2cm of x1] {};
\node (dis2) [right =.3cm of temdis11] {};
\node (dis1) [below =-.1cm of dis2]{\scriptsize$\Delta$};
\draw[-,thin,black,dotted](x1)edge node[swap,near start]{}  (y1);

\draw[-,thick,black,dashed](x1)edge node[swap,near start]{}  (x3);
\draw[-,thick,black,dashed](y1)edge node[swap,near start]{}  (y3);

\path (y3.east |-x3.north)+(.2,+.35) node(cor1){};
\path (x2.west|-x1.south)+(-.45,-.35) node(cor2){};
\path[
draw=black!100,thick]
(cor1) rectangle  (cor2);
\end{tikzpicture}
	}}}\hspace{-.3cm}\,
	\subfloat[]
	{\label{Fig:Ex_Biggerset_n}
		\centering{
			\scalebox{.9}
			{\centering
				\pgfdeclarelayer{background} \pgfdeclarelayer{foreground}
\pgfsetlayers{background,main,foreground}
\par
\par
\tikzset{cross/.style={cross out, draw=black, minimum size=2*(#1-\pgflinewidth), inner sep=0pt, outer sep=0pt},
	cross/.default={3pt}}
\begin{tikzpicture}[]

\draw (0,0) node[circle,draw=black, fill=black, inner sep=0pt,minimum size=2.5pt](x1){};
\draw (0,-.2) node {$x_1$};

\draw (0,.4) node[circle,draw=black, fill=black, inner sep=0pt,minimum size=2.5pt](x2) {};
\draw (-0.25,.4) node {$x_2$};

\draw (0,1) node[circle,draw=black, fill=black, inner sep=0pt,minimum size=2.5pt](xi) {};
\draw (-0.25,1) node {$x_i$};

\draw (0,1.68) node[circle,draw=black, fill=black, inner sep=0pt,minimum size=2.5pt](xn-1) {};
\draw (0,1.68) node {};

\draw (0,2) node[circle,draw=black, fill=black, inner sep=0pt,minimum size=2.5pt](xn) {};
\draw (0,2.15) node {$x_n$};

\draw (1.1,0) node[circle,draw=black, fill=black, inner sep=0pt,minimum size=2.5pt](y1){};
\draw (1.1,-.2) node {$y_1$};

\draw (1.1,.4) node[circle,draw=black, fill=black, inner sep=0pt,minimum size=2.5pt](y2){};
\draw (.85,.4) node {$y_2$};

\draw (1.1,1) node[circle,draw=black, fill=black, inner sep=0pt,minimum size=2.5pt](yi){};
\draw (.85,1) node {$y_i$};

\draw (1.1,1.68) node[circle,draw=black, fill=black, inner sep=0pt,minimum size=2.5pt](yn-1) {};
\draw (1.1,1.68) node {};

\draw (1.1,2) node[circle,draw=black, fill=black, inner sep=0pt,minimum size=2.5pt](yn){};
\draw (1.1,2.15) node {$y_n$};

\node(temdis11)  [above =.2cm of x1] {};
\node (dis2) [right =.3cm of temdis11] {};
\node (dis1) [below =-.1cm of dis2]{\scriptsize$\Delta$};
\draw[-,thin,black,dotted](x1)edge node[swap,near start]{}  (y1);

\draw[-,thick,black,dashed](x1)edge node[swap,near start]{}  (xn);
\draw[-,thick,black,dashed](y1)edge node[swap,near start]{}  (yn);

\path (yn.east |-xn.north)+(.2,+.35) node(cor1){};
\path (xi.west|-x1.south)+(-.45,-.35) node(cor2){};
\path[
draw=black!100,thick]
(cor1) rectangle  (cor2);
\end{tikzpicture}
	}}}
\caption{The Euclidean distance between two line segments $l_1$ and $l_2$ is $\Delta$ which is shown in Figure \ref{Fig:Ex_Biggerset_0}. If these two line segments are discretized into two sets of two points each as in Figure \ref{Fig:Ex_Biggerset_1}; two sets of three points each as in Figure \ref{Fig:Ex_Biggerset_2}; and two sets of $n$ points each as in Figure \ref{Fig:Ex_Biggerset_n}, then the GOSPA metric is $2\Delta$, $3\Delta$ and $n\Delta$ respectively. The OSPA and COSPA metrics are the same and equal to $\Delta$ for all cases.
}
	\label{Fig:Ex_Biggerset}
\end{figure}
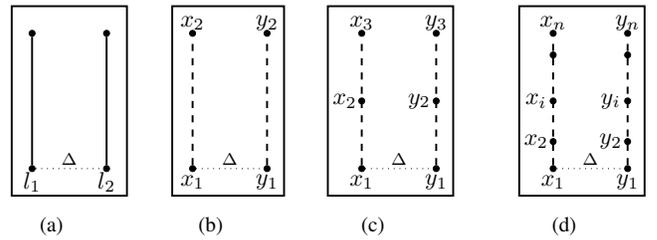

The normalization of the distance between two finite sets plays an important role for measuring how close these two finite sets are because it scales the total error, which is the minimum sum of all distances of pairs of vectors and all distances of unpaired vectors, to be within the interval $[0,c]$ for the OSPA metric and to  be within the half-closed interval $[0,\dot{c}+\dot{c}\xi)$ where $0\le\xi\le 1$  for the COSPA metric. Without the normalization, the GOSPA metric is simply the sum. Hence, if one/both of these sets are large and distinguishable, the total error of these two sets is large and hence the GOSPA distance is large. Considering Figures \ref{Fig:Ex_RemDe_1} and \ref{Fig:Ex_RemDe_3}, the GOSPA metric concludes that $Y_d$ is worse than $Y_b$ because their distances to the same set $X$ gives the bigger value to the first compared to the second. This is summarized in Table \ref{table:Fig:Ex_RemDe1}. Clearly, the unnormalized OSPA distance is not a proper distance measure because it is proportional to the size of the bigger set. Therefore, the GOSPA metric may not be a suitable tool to measure how close two finite sets of vectors are. 
Now consider Figure \ref{Fig:Ex_Biggerset} as an example. There are two  parallel line segments in Figure \ref{Fig:Ex_Biggerset_0} and the distance between these line segments is $\Delta$. If these two line segments are discretized into two sets of two points each, with $X=\{x_1,x_2\}$ and $Y=\{y_1,y_2\}$ as in Figure \ref{Fig:Ex_Biggerset_1}, the GOSPA distance will be $\Delta\sqrt[p]{2}$. Similarly, if these two lines are discretized  into two sets of $n$ points each, $X=\{x_1,x_2,\ldots,x_n\}$ and $Y=\{y_1,y_2,\ldots,y_n\}$ as in Figure \ref{Fig:Ex_Biggerset_n}, then the GOSPA metric will be $\Delta\sqrt[p]{n}$. This means the distance for the scenario in Figure \ref{Fig:Ex_Biggerset_n} is larger than the distance for the scenario in Figure \ref{Fig:Ex_Biggerset_1}. This is not intuitively reasonable because all scenarios in Figure \ref{Fig:Ex_Biggerset} have the same distance, $\Delta$, between the two line segments $l_1$ and $l_2$ in Figure \ref{Fig:Ex_Biggerset_0}. OSPA and COSPA give the same distance $\Delta$ for all scenarios in Figure \ref{Fig:Ex_Biggerset}. The explanation and computation of the three metrics for Figures \ref{Fig:Ex_Biggerset_1}-\ref{Fig:Ex_Biggerset_n} are summarized in Table \ref{table:Fig:Ex_Biggerset}. 
\renewcommand{\arraystretch}{1.7}
\begin{table}[htpb!]%
	\caption{\textbf{Analysis of the three metrics 
			for Figures \ref{Fig:Ex_RemDe_1} - \ref{Fig:Ex_RemDe_3}}}
	\begin{center}
		\begin{tabular}
			{|l||l|l|l|}\hline
			\!Metric\!&\!Fig.\ref{Fig:Ex_RemDe_1}  ($Y_b$)&\!Fig.\ref{Fig:Ex_RemDe_3} ($Y_d$)&Which one 
			 is closer to $X$?\\ \hline\hline
			\!OSPA & \!$\frac{c+\Delta}{2}$ &$\frac{2\Delta+c}{3}$&$Y_d$\\ \hline
			GOSPA& $\Delta+\frac{c}{\alpha}$&\!$2\Delta+\frac{c}{\alpha}$&$Y_b$\\ \hline
			\!COSPA\!& \!$\frac{\dot{c}+\Delta}{2}$ &\!$\frac{2\Delta+\dot{c}}{3}$&$Y_d$\\ \hline
			Intuition& & &$Y_d$\\ \hline
		\end{tabular}
		\label{table:Fig:Ex_RemDe1}
	\end{center}
\end{table}
\renewcommand{\arraystretch}{1}
\renewcommand{\arraystretch}{1.7}
\begin{table}[htpb!]%
	\caption{\textbf{Analysis of the three metrics 
			 for Figures \ref{Fig:Ex_Biggerset_1}-\ref{Fig:Ex_Biggerset_n}}}
	\begin{center}
		\begin{tabular}
			{|l||l|l|l|l|}\hline
			Metric&Fig.\ref{Fig:Ex_Biggerset_1}&Fig.\ref{Fig:Ex_Biggerset_2}&Fig.\ref{Fig:Ex_Biggerset_n}&Are the distances 
			 the same?\\ \hline\hline
			OSPA &$\Delta$& $\Delta$ &$\Delta$&Yes\\ \hline
			GOSPA&$2\Delta$& $3\Delta$&$n\Delta$&No\\\hline
			COSPA&$\Delta$& $\Delta$ &$\Delta$&Yes\\ \hline
			Intuition&& &&Yes\\ \hline
		\end{tabular}
		\label{table:Fig:Ex_Biggerset}
	\end{center}
\end{table}
\renewcommand{\arraystretch}{1}

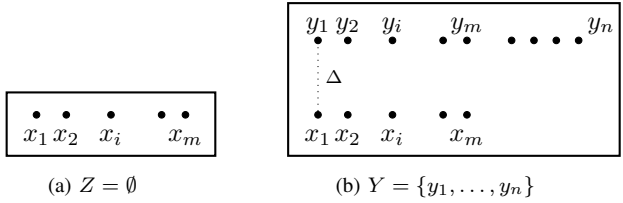
\begin{figure}[htbp!]
	\hspace{-.7cm}%
	\centering\subfloat[$Z=\emptyset$]
	{\label{Fig:Ex_Biggersetn_0}
		\centering{
			\scalebox{.99}
			{\centering
				\pgfdeclarelayer{background} \pgfdeclarelayer{foreground}
\pgfsetlayers{background,main,foreground}
\par
\par
\tikzset{cross/.style={cross out, draw=black, minimum size=2*(#1-\pgflinewidth), inner sep=0pt, outer sep=0pt},
	cross/.default={3pt}}
\begin{tikzpicture}[]

\draw (0,0) node[circle,draw=black, fill=black, inner sep=0pt,minimum size=2.5pt](x1){};
\draw (0,-.3) node {$x_1$};

\draw (0.4,0) node[circle,draw=black, fill=black, inner sep=0pt,minimum size=2.5pt](x2) {};
\draw (0.4,-.3)node {$x_2$};

\draw (1,0) node[circle,draw=black, fill=black, inner sep=0pt,minimum size=2.5pt](xi) {};
\draw (1,-.3) node {$x_i$};

\draw (1.68,0) node[circle,draw=black, fill=black, inner sep=0pt,minimum size=2.5pt](xn-1) {};

\draw (2,0) node[circle,draw=black, fill=black, inner sep=0pt,minimum size=2.5pt](xn) {};
\draw (2,-.3) node {$x_m$};
\path (xn.east |-x1.north)+(.35,+.25) node(cor1){};
\path (x1.west|-x1.south)+(-.35,-.5) node(cor2){};
\path[
draw=black!100,thick]
(cor1) rectangle  (cor2);
\end{tikzpicture}
	}}}\hspace{-.3cm}\,
	\subfloat[$Y=\{y_1,\ldots,y_n\}$]
	{\label{Fig:Ex_Biggersetn_2n}
		\centering{
			\scalebox{.99}
			{\centering
				\pgfdeclarelayer{background} \pgfdeclarelayer{foreground}
\pgfsetlayers{background,main,foreground}
\par
\par
\tikzset{cross/.style={cross out, draw=black, minimum size=2*(#1-\pgflinewidth), inner sep=0pt, outer sep=0pt},
	cross/.default={3pt}}
\begin{tikzpicture}[]

\draw (0,0) node[circle,draw=black, fill=black, inner sep=0pt,minimum size=2.5pt](x1){};
\draw (0,-.3) node {$x_1$};

\draw (0.4,0) node[circle,draw=black, fill=black, inner sep=0pt,minimum size=2.5pt](x2) {};
\draw (0.4,-.3)node {$x_2$};

\draw (1,0) node[circle,draw=black, fill=black, inner sep=0pt,minimum size=2.5pt](xi) {};
\draw (1,-.3) node {$x_i$};

\draw (1.68,0) node[circle,draw=black, fill=black, inner sep=0pt,minimum size=2.5pt](xn-1) {};

\draw (2,0) node[circle,draw=black, fill=black, inner sep=0pt,minimum size=2.5pt](xn) {};
\draw (2,-.3) node {$x_m$};

\draw (0,1) node[circle,draw=black, fill=black, inner sep=0pt,minimum size=2.5pt](y1){};
\draw (0,1.2) node {$y_1$};

\draw (.4,1) node[circle,draw=black, fill=black, inner sep=0pt,minimum size=2.5pt](y2){};
\draw (.4,1.2) node {$y_2$};

\draw (1,1) node[circle,draw=black, fill=black, inner sep=0pt,minimum size=2.5pt](yi){};
\draw (1,1.2) node {$y_i$};

\draw (1.68,1) node[circle,draw=black, fill=black, inner sep=0pt,minimum size=2.5pt](yn-1) {};

\draw (2,1) node[circle,draw=black, fill=black, inner sep=0pt,minimum size=2.5pt](yn){};
\draw (2,1.2) node {$y_m$};

\draw (2.6,1) node[circle,draw=black, fill=black, inner sep=0pt,minimum size=2.5pt](yn1){};

\draw (2.9,1) node[circle,draw=black, fill=black, inner sep=0pt,minimum size=2.5pt](yn2){};

\draw (3.2,1) node[circle,draw=black, fill=black, inner sep=0pt,minimum size=2.5pt](yn2){};

\draw (3.5,1) node[circle,draw=black, fill=black, inner sep=0pt,minimum size=2.5pt](y2n1){};
\draw (3.8,1.2) node {$y_n$};

\node(temdis11)  [above =.33cm of x1] {};
\node (dis2) [right =-.15cm of temdis11]{\scriptsize$\Delta$};
\draw[-,thin,black,dotted](x1)edge node[swap,near start]{}  (y1);


\path (y2n1.east |-y2n1.north)+(.5,+.45) node(cor1){};
\path (x1.west|-x1.south)+(-.35,-.5) node(cor2){};
\path[
draw=black!100,thick]
(cor1) rectangle  (cor2);
\end{tikzpicture}
	}}}
	\caption{The ground truth is $X=\{x_1,\ldots,x_m\}$. In Figure \ref{Fig:Ex_Biggersetn_0}, an algorithm estimates no target, i.e. $Z=\emptyset$ while in Figure \ref{Fig:Ex_Biggersetn_2n} an algorithm estimates $n$ targets ($n>m$), i.e $Y=\{y_1,y_2,\ldots,y_{n}\}$ where $\Delta<c$.}
	\label{Fig:Ex_2Biggersetn}
\end{figure}

Furthermore, the absence of normalization makes GOSPA inconsistent when comparing the empty set with a non-empty set using another non-empty set as a reference. Take Figure \ref{Fig:Ex_2Biggersetn} as an example. The computation and the comparison of the results (of the three metrics) against intuitive thinking are summarized in Table \ref{table:Ex_2Biggersetn}.

\renewcommand{\arraystretch}{1.7}
\begin{center}
	\begin{table}[htpb!]%
		\caption{\textbf{Analysis of the three metrics 
				for Figure \ref{Fig:Ex_2Biggersetn}}}
		\begin{center}
			\begin{tabular}
				{|l||l|l|l|}\hline
				Metric\!&Fig.\ref{Fig:Ex_Biggersetn_0} ($Z$)&Fig. \ref{Fig:Ex_Biggersetn_2n} ($Y$)& Which one 
				is closer to $X$?\\ \hline\hline
				OSPA &$c$ & $\frac{m\Delta+(n-m)c}{n}$& $Y$ \\ \hline
				GOSPA&$c\frac{m}{\alpha}$& $m\Delta+\frac{(n-m)c}{\alpha}$
				& $Y$ if $\boldsymbol{\Delta< c\frac{2m-n}{m\alpha}}$
				\\\hline
				COSPA&$\dot{c}\left(2-\frac{1}{n}\right)$ & $\frac{m\Delta+(n-m)\dot{c}}{n}$&  $Y$\\ \hline
				Intuition&&&$Y$\\ \hline
			\end{tabular}
			\label{table:Ex_2Biggersetn}
		\end{center}
	\end{table}
\end{center}
\renewcommand{\arraystretch}{1}
\section{Experiment}\label{S:Experiment}
\begin{figure}[htbp!]
	\includegraphics[width=\textwidth/2]
	{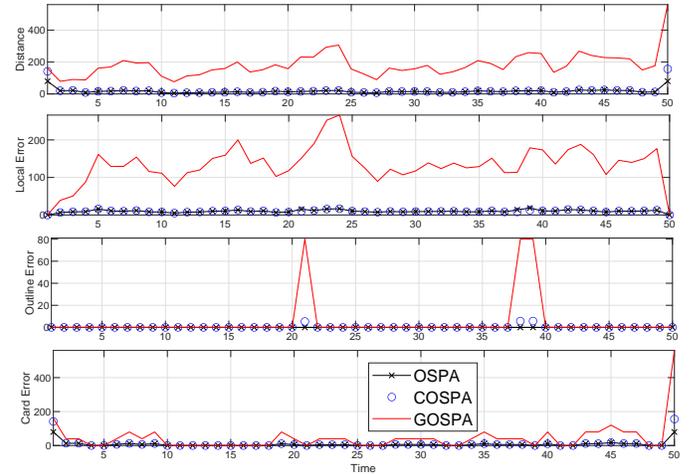}
	\caption{Error versus time calculated of OSPA, GOSPA and COSPA metric.}
	\label{Fig_All3Metrics}
\end{figure}

\begin{figure}[htbp!]
	\includegraphics[width=\textwidth/2]
	{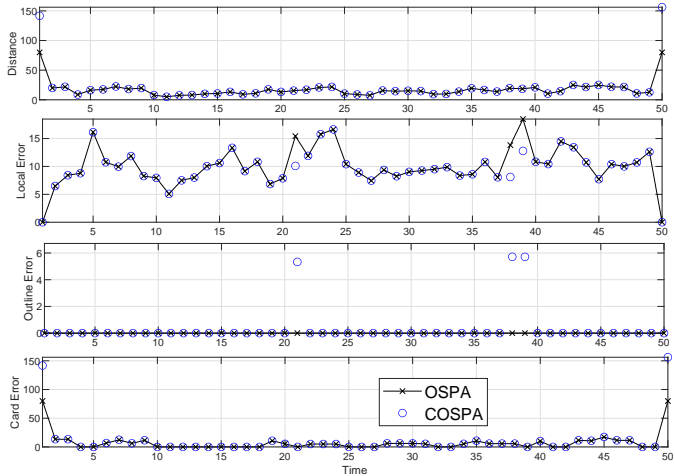}
	\caption{Error versus time calculated under OSPA and COSPA metric.}
	\label{Fig:OSPA_COSPATime}
\end{figure}

\begin{figure}[htbp!]
	\includegraphics[width=\textwidth/2]
	{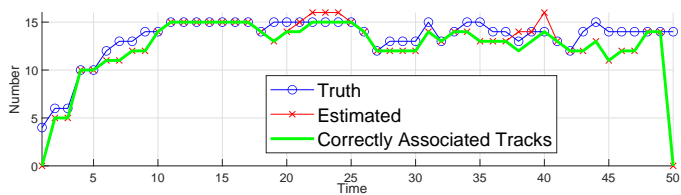}
	\caption{Target numbers are plotted versus time}
	\label{Fig_TruthEstimateAssociateTrack}
\end{figure}

\begin{figure}[htbp!]
	\includegraphics[width=\textwidth/2]
	{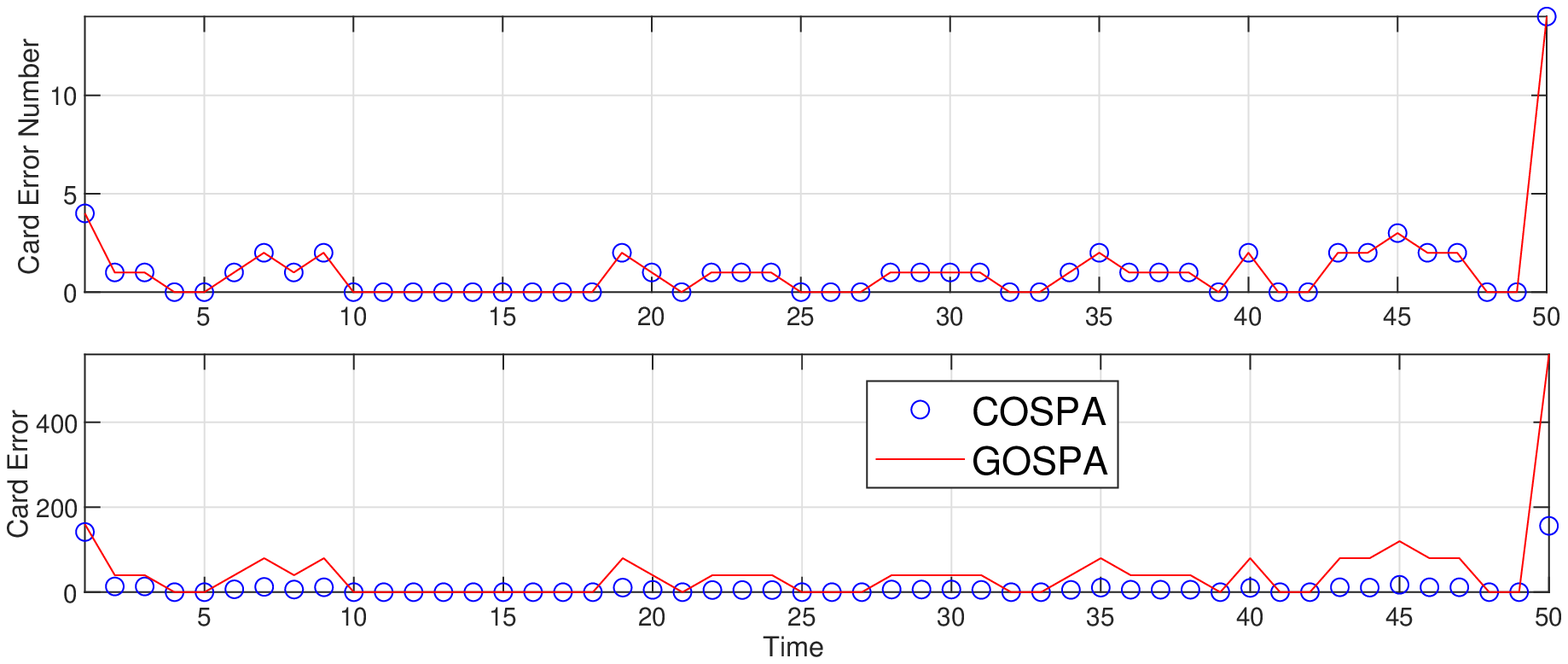}
	\caption{Difference in number between two sets and Cardinality Error versus time calculated under GOSPA and COSPA metrics.}
	\label{Fig:CardErrorCardNoGOSPACOSPA}
\end{figure}

\begin{figure}[htbp!]
	\includegraphics[width=\textwidth/2]
	{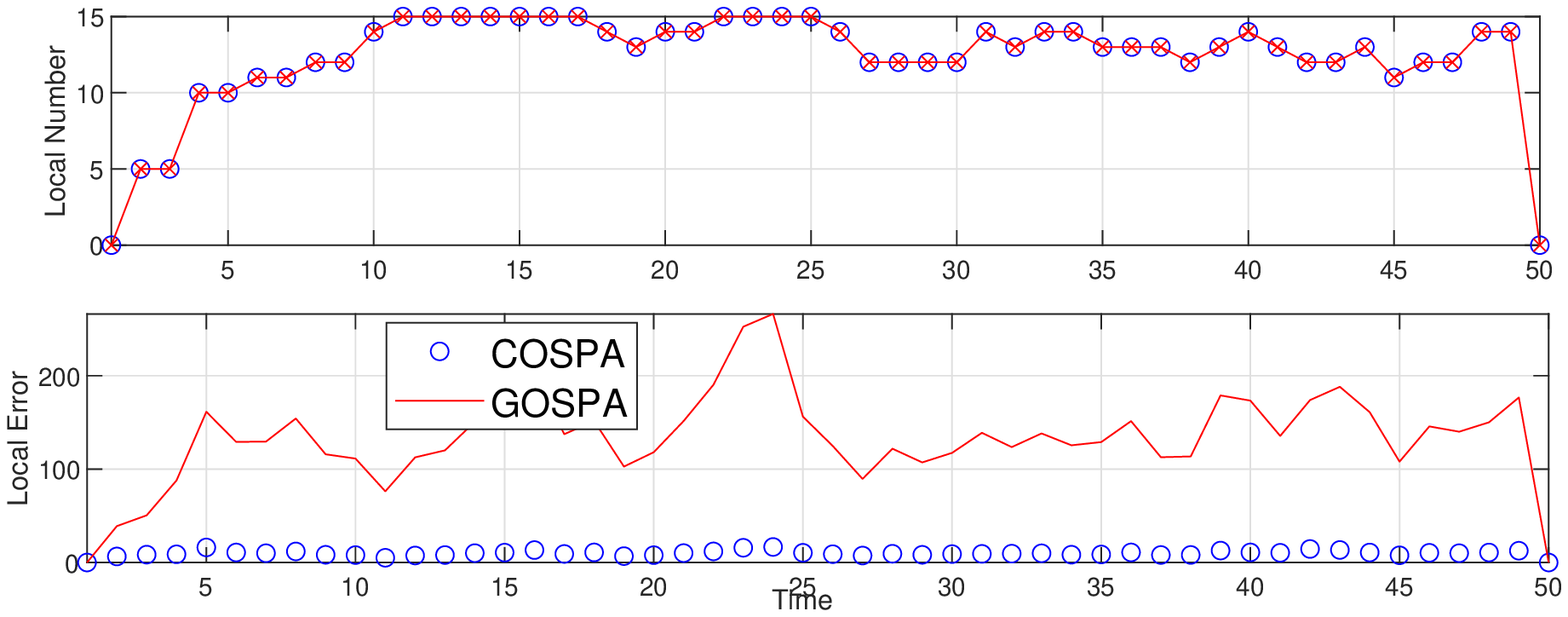}
	\caption{Difference in number between two sets and Cardinality Error versus time calculated under GOSPA and COSPA metrics.}
	\label{Fig:LocalAndNumerGOSPACOSPA}
\end{figure}

We demonstrate the proposed metric by evaluating an multi-target tracking (MTT) algorithm together with OSPA and GOSPA metrics. We use the data and one of the result produced from MTT algorithm in \cite{TuyetPMMHT14} $38$ targets move from top right or middle of the surveillance
area to bottom left, and middle of the surveillance area to top
right. Each target 
survives with probability $99\%$ and is detected with $80\%$. The measurements are added noize with  zero mean Gaussian process. The detected measurements are immersed in clutter modeled as a Poisson RFS with the average number of clutter returns per unit volume is $50$. 
In this example, we use the OSPA metric \cite{Schuhmacher2008}, the GOSPA metric \cite{rah2016GOSPA} and the proposed COSPA metric to compute the distances between the truth tracks and the estimated tracks (produced by the MTT algorithm).
\begin{figure}[htbp!]
	\centering
	\includegraphics[width=\textwidth/3+1cm]
	{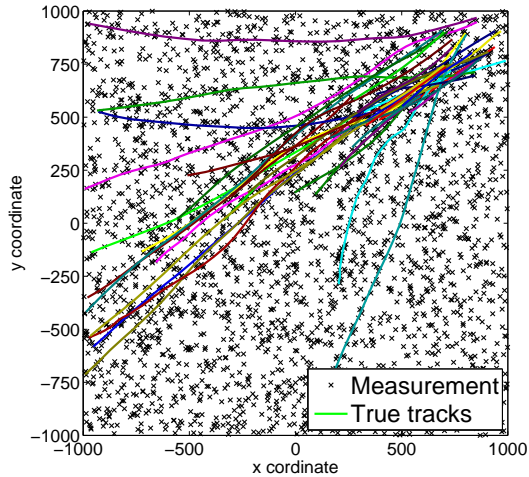}
	\caption{Ground truth targets are immersed with their measurements and clutter.}
	\label{FigGrdtruth}
\end{figure}
For computation of the three metrics, we chose the cut-off parameter for the three metrics $c=80$, $\alpha=2$, $\xi=1$, $p=1$ and $\dot{c}=c+1$. 
In this scenario,  Figure \ref{Fig_All3Metrics} shows that at any time $t$ ($t=1,\ldots,50$) the GOSPA metric has much bigger error compared to OSPA and COSPA because GOSPA is a sum of all spatial distances and cardinality error at that time while OSPA and COSPA are the average of this sum. While Figure \ref{Fig:OSPA_COSPATime} shows that most of the time OSPA and COSPA have the same errors except at time where the cut-off is applied or one of the two sets but not both is empty. It is because the cut-off distance $c$ and the penalty to each cardinality error are different in COSPA but the same in OSPA. Figure \ref{Fig_TruthEstimateAssociateTrack} shows that most of the time all objects in the smaller sets associate correctly with objects in a bigger set except at time $t=21,38,39$ (two objects are correctly associated if their distance is smaller than the cut-off $c$). Indeed, OSPA gives value as a cut-off $c$ to the distance between the two sets at time $t=1$ and $t=50$ at which no track is detected. Furthermore, the distances and cardinality error at time steps $t=21,38$ and $t=39$ are the same in both COSPA and OSPA but OSPA local error is the sum of COSPA local error and COSPA outline error. It is because the distance between two vectors not to be smaller than cut-off parameter $c$ is considered as the wrongly associated tracks in COSPA. It happens because in COSPA the cut-off parameter $c$ is smaller than the penalty for each cardinality error $\dot{c}$.
Figure \ref{Fig:CardErrorCardNoGOSPACOSPA} shows that the GOSPA cardinality error is proportional to the difference in number between bigger set and the smaller set.
Indeed, the graph of cardinality error is $c/\alpha=80/2$ times the graph of the cardinality number. Similarly, Figure \ref{Fig:LocalAndNumerGOSPACOSPA} shows that the GOSPA localization error is proportional to the number of spatial distance across two sets.

\section{Conclusion}\label{S:Conclusion}
This paper has discussed limitations of the  OSPA metric via some simple scenarios. Furthermore, some certain limitations of the GOSPA metric were discussed. GOSPA tries to overcome the limitations of the OSPA metric by removing the normalization of the OSPA metric and setting the cardinality penalty different from the cut-off parameter by multiplying the cut-off parameter with a positive number $\frac{1}{\alpha}$  that is larger than or equal to $\frac{1}{2}$ ($0<\alpha\le 2$). This alteration of the OSPA metric results in a greater penalty for a distance $\Delta$ between two vectors 
than the penalty for each cardinality penalty if $c>\Delta>c/\alpha$. Hence the GOSPA metric will normally favor the empty set over the non-empty set when $\alpha>1$ and these two sets are compared with another non-empty set of the same size as the non-empty set and the distances between pairs of vectors across the two non-empty sets are equal to or larger than the cardinality penalty; or the localization error can be penalized more than the cardinality penalty if a distance $\Delta$ between two vectors is larger than $c/\alpha$ ($\Delta<c$). 
Furthermore, the lack of normalization in the GOSPA metric makes it unreliable for measuring the distance between two non-empty sets. 

The proposed COSPA metric was developed to overcome the shortcomings mentioned by \cite{Schuhmacher2008} and also provide a practical assessment of the MTF  or multiple target tracking algorithms at a particular time in terms of missing targets, false targets, incorrectly associated targets and correctly associated target. Furthermore, the identities of missing targets, false targets and pairs of associated targets are provided in the process of calculating this metric (available as Matlab code in \cite{TuyetCodeCOSPA}). The COSPA metric retains the advantages of the OSPA metric, unlike the GOSPA metric. Thorough analysis of the COSPA metric reveals no major weaknesses, 
noting that the penalty for each extra element in two sets is always larger than or equal to the cut-off distance between two vectors. The choice of the cut-off is problem dependent. We analysed the proposed metric together with other two metrics with some simple scenarios which shows the consistency and improvement compared the other two metric OSPA and GOSPA. We also use the set of tracks resulting from a multiple target tracking algorithm to evaluate the proposed metric compared with other two metrics.
\section*{Acknowledgement}
The author would like to thank Professor Henk Blom for the discussions, suggestions and comments.
\bibliographystyle{IEEEtran}
\bibliography{references_new}
\end{document}